\documentclass{pic2012}
\usepackage{lineno}
\usepackage{url}
\usepackage[centertags]{amsmath}
\usepackage{ulem}
\usepackage{textcomp}
\usepackage{graphicx}
\usepackage{amsfonts}
\usepackage{amssymb}
\usepackage{amsthm}
\usepackage{newlfont}
\usepackage[]{subfigure}
\usepackage{graphicx}

\def\runauthor{E. Stanecka, on behalf of the ATLAS Collaboration}
\def\shorttitle{The ATLAS Inner Detector operation and tracking performance.}

\begin{document}
\title{The ATLAS Inner Detector operation, data quality
and tracking performance. }

\author{E.Stanecka, on behalf of the ATLAS Collaboration}

\address{Institute of Nuclear Physics PAN\\
ul. Radzikowskiego 152\\
31-342 Krakow, Poland\\
E-mail: ewa.stanecka@ifj.edu.pl}

\maketitle

\abstracts{ The ATLAS Inner Detector is responsible for particle
tracking in ATLAS experiment at CERN Large Hadron Collider (LHC) and
comprises silicon and gas based detectors. The combination of both
silicon and gas based detectors provides high precision impact
parameter and momentum measurement of charged particles, with high
efficiency and small fake rate. The ID has been used to exploit
fully the physics potential of the LHC since the first proton-proton
collisions at 7 TeV were delivered in 2009. The performance of track
and vertex reconstruction is presented, as well as the operation
aspects of the Inner Detector and the data quality during the many
months of data taking. }

\section{Introduction}
\begin{figure}[h]
\begin{centering}
\begin{minipage}[b]{75mm}
\centering
\includegraphics[width=75mm]{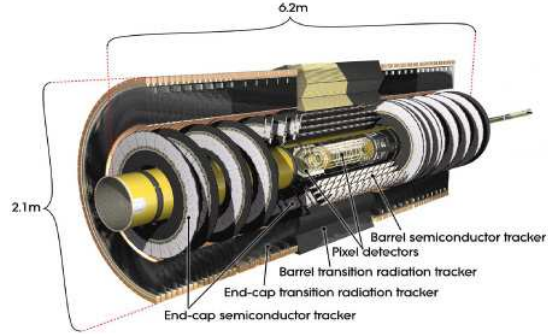}\\
\end{minipage}
  \caption{Geometrical layout of ATLAS Inner Detector.
  \label{fig:id}}
\end{centering}
\end{figure}

The Inner Detector (ID) \cite{id} forms the tracking system for ATLAS
experiment \cite{atlas}. It is designed to reconstruct charged particle tracks
and vertices, to measure the momentum of charged particles above a
given transverse momentum  ($p_{T}$) threshold
and within the pseudorapidity range $|\eta| < 2.5$. The ID operates
in a high fluence environment and to achieve required momentum and
vertex resolution the detector granularity must be very fine. The ID
consists of three types of tracking components (from innermost
layer): Pixel Detector \cite{pixel}, SemiConductor Tracker (SCT) \cite{sct} and Transition
Radiation Tracker (TRT) \cite{trt}. The three sub-detectors are contained
within a cylindrical envelope of length § 3512 mm and of radius 1150
mm, surrounded by a solenoid providing magnetic field of 2 T. Figure
\ref{fig:id} illustrates the geometrical layout of the Inner
Detector.

The Pixel Detector is situated the closest to the interaction point
and has the highest granularity. In total there are about 80 million
readout channels in the whole Pixel Detector. The intrinsic spatial
resolution of individual Pixel Detector modules, is 10 $\mu$m in $R\phi$ and
115 $\mu$m in $z$. The SCT is a silicon detector with microstrips
which surrounds the Pixel Detector layers. As track density decreases at
larger radii, microstrip detectors can have fewer read-out channels.
It is designed to provide 8 measurements per track with resolution
of 16 $\mu$m in $R\phi$ and 580 $\mu$m in $z$. The TRT is the most
outer part of ID. It provides a large number of tracking
measurements (typically $>$ 30 hits per track), good pattern
recognition and it contributes to particle identification. The TRT
is a light-weight detector, composed of gaseous proportional
counters (straws) embedded in a radiator material. The straws are
filled with 70\% Xe, 27\% CO$_{2}$ and 3\% O$_{2}$ gas mixture with
5-10 mbar over-pressure, for which the operational drift radius
accuracy is $\sim$130 $\mu$m.

\section{Detector operations, data taking and data quality}

LHC delivered an integrated luminosity of 5.6 fb$^{-1}$ of
proton-proton collision data at the center-of-mass energy of 7 TeV in
2010 and 2011. In 2012 the center-of-mass energy was increased to 8
TeV and the LHC luminosity was upgraded significantly. ATLAS
recorded integrated luminosity of impressive value of 14.3 fb$^{-1}$
by October 2012 \cite{lumi}. The ID has been fully operational
throughout all data taking periods and delivered excellent
data taking performance. 
Luminosity weighted relative fraction of good quality data delivered
during 2012 stable beams in pp collisions by the Pixel Detector, SCT, and TRT
subsystems  was 99.9\%, 99,4\% and 99,8\% accordingly
\cite{quality}.

Such an excellent data taking performance was possible thanks to
reliable and robust Data Acquisition Systems (DAQ) and Detector
Control System (DCS). The Data Acquisition Systems of all ID
sub-detectors have proved to be highly reliable with excellent data
taking efficiency. During operation several enhancements were
introduced into DAQ in order to avoid potential sources of
inefficiency. One of the possible inefficiency sources are readout
chip errors (e.g. spontaneous corruption of module configuration)
caused by Single Event Upsets (SEUs). SCT introduced online
monitoring of chip errors in the data and the automatic
reconfiguration of the modules with errors. In addition, an
automatic global reconfiguration of all SCT module chips every
$\sim$30 minutes was implemented, as a precaution against subtle
deterioration in chip configurations as a result of SEUs
\cite{dave}.
 The main component in the DAQ is the Readout Driver
Board (ROD), which provides the front-end data flow, data processing
and control of the detector modules. If any ROD experiences an error
condition, it will exert a BUSY signal to stop ATLAS data taking. In
2010, each subsystem implemented an automatic removal of a busy ROD
from the ATLAS readout, thereby enabling ATLAS to continue data
taking while the cause of the BUSY was corrected. Automatic
re-integration of a recovered ROD was also implemented. An automatic
re-synchronisation procedure of the TRT RODs was also introduced.
This procedure is invoked during the LHC ramp and whenever
synchronization is lost. It allows for the TRT to continue taking
data without stopping and restarting the run.

The Detector Control System (DCS) supervises, besides the individual subsystems, ID detector components,
provides information about conditions inside the detector, assures
optimal working conditions and provides protection mechanisms. The
main ID DCS subsystems are: Evaporating Cooling to keep silicon
detector cooled to 
$- 10\,^{\circ}\mathrm{C}$, Heater Pad systems that ensures thermal
shield between silicon detectors and TRT operating in room
temperature. There is also set of projects dedicated to beam
condition monitoring and radiation doses monitoring inside ID
volume.
To protect the front-end electronics from the potentially harmful
effects of beam incidents, the HV was only switched fully on for the
Pixel Detector and SCT when stable beam conditions were declared.
Outside of stable beams, the Pixel Detector HV was off, and the SCT
was operated with reduced HV. In SCT and Pixel Detector DCS the
automatic turn-on, so called "warm start", was implemented in order
to maximize the time of data taking. Detectors were set
automatically to ready-for-data-taking state immediately after
stable beams were declared by LHC and beam parameters, measured by
ID beam condition monitoring system, were correct. Typical time to
ready for data-taking in silicon detectors was $\sim$1 minute.

\section{Radiation Damage}
\begin{figure}[h]
\begin{centering}
\begin{minipage}[b]{60mm}
\centering
\includegraphics[width=60mm]{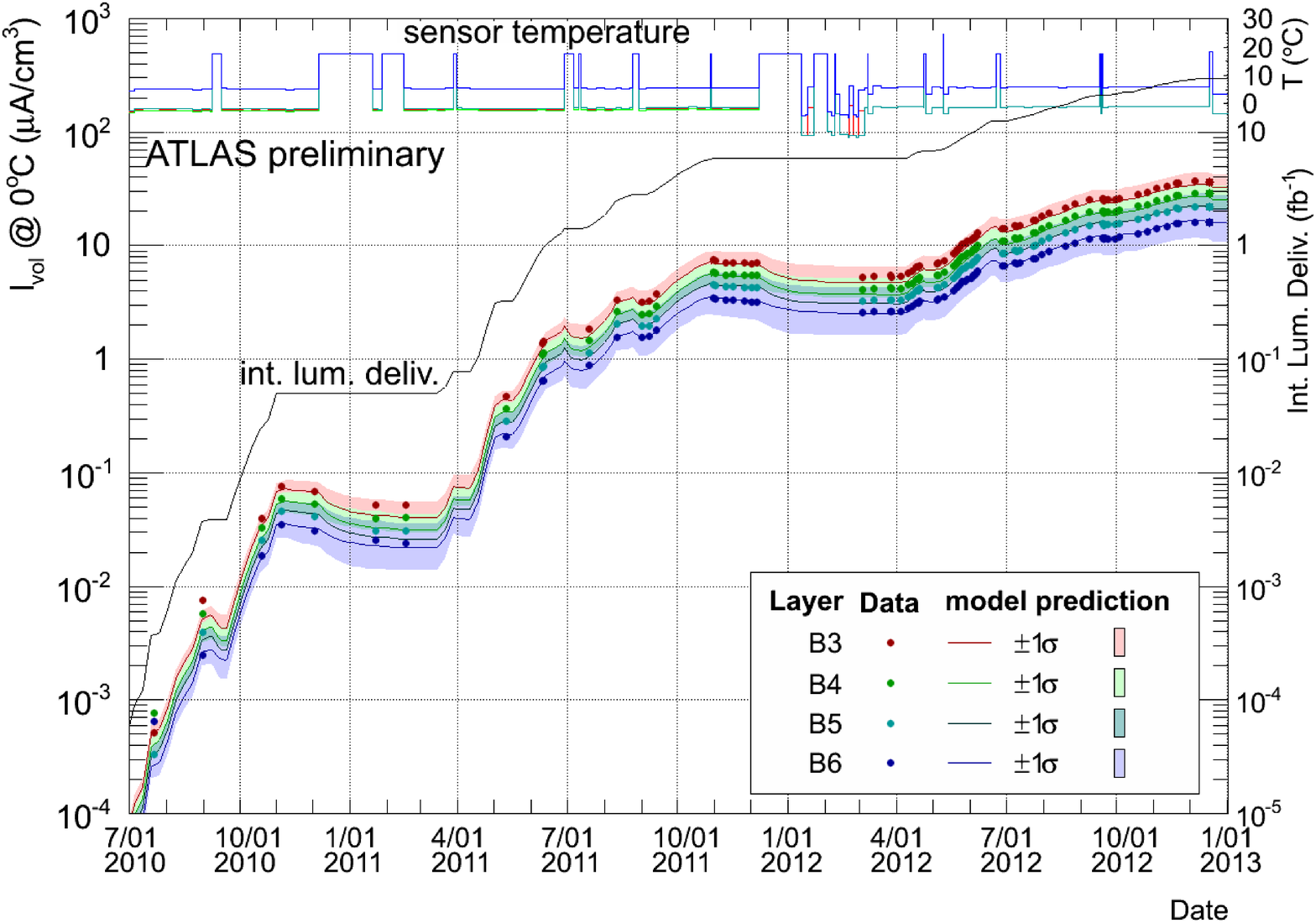}\\
\end{minipage}
\begin{minipage}[b]{60mm}
\centering
\includegraphics[width=60mm]{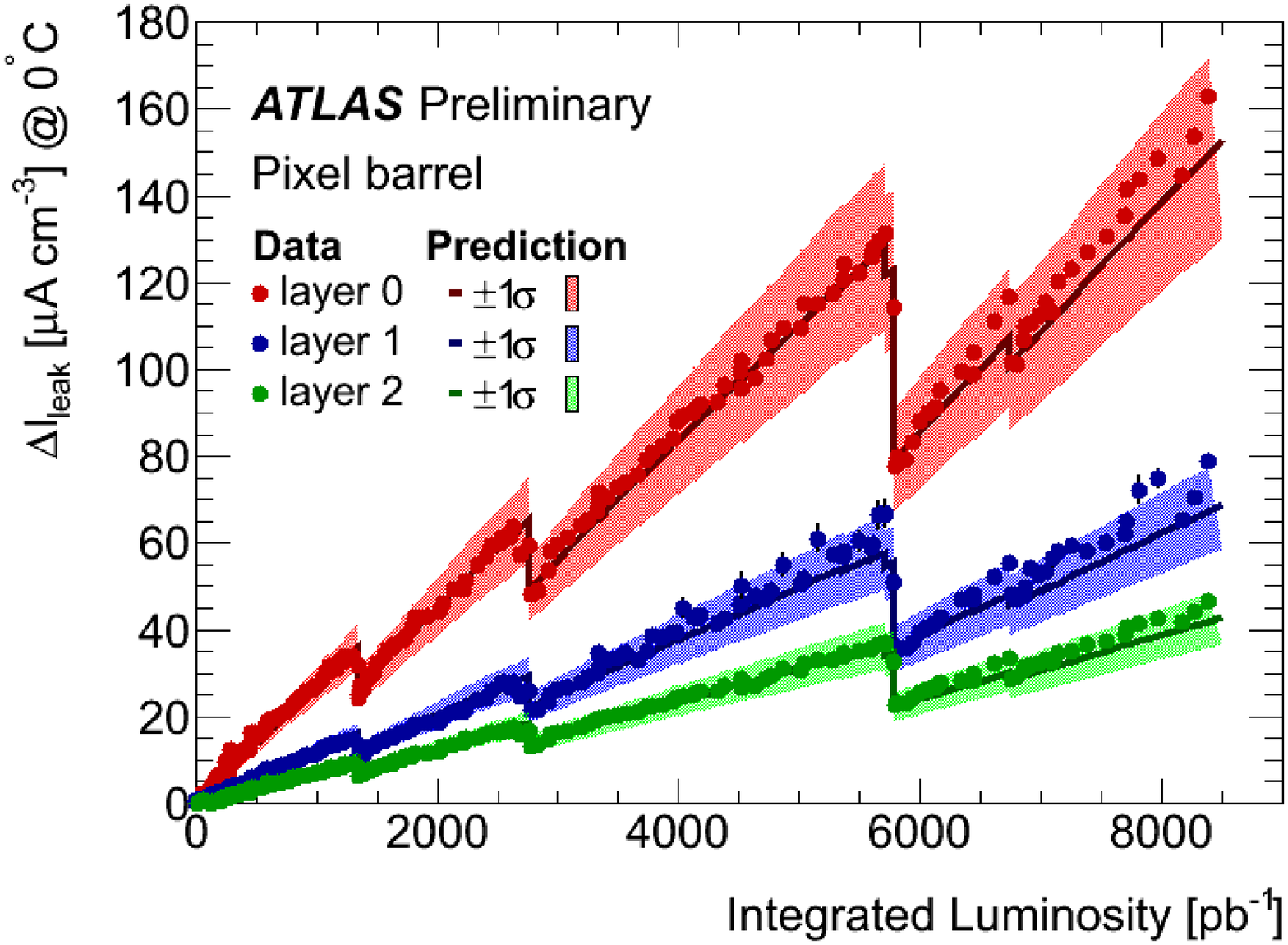}\\
\end{minipage}
  \caption{SCT barrel leakage currents during 2010, 2011 and 2012, showing correlations
with delivered luminosity and temperature, and compared to
predictions from Monte Carlo (left) \protect\cite{sct_pub}. The
averaged reverse-bias current for all Pixel Detector modules in the different
Barrel layers as a function of the integrated luminosity. The model
predictions underestimate the data, thus have been scaled up by 15\%
(Layer 0) or 25\% (Layer 1 and 2), respectively (right) \protect\cite{pix_pub}.
  \label{fig:leak}}
\end{centering}
\end{figure}

Radiation damage effects in SCT and Pixel Detector became visible in 2011 and
they are increasing with luminosity and time. Significant increases
in leakage currents have been observed in silicon detectors, as
expected from bulk damage due to non-ionising radiation. Figure
\ref{fig:leak} shows SCT barrel leakage currents during 2010, 2011
and 2012, and evolution of high voltage current for Pixel Detector. The
leakage current increase correlates closely with delivered
luminosity and temperature cycles. This dependence is
well-understood and agrees with leakage current predictions derived
from the temperature profiles and fluence from FLUKA \cite{fluka}
simulations.

\section{Track and vertex reconstruction performance}
\begin{figure}[h]
\begin{centering}
\begin{minipage}[b]{60mm}
\centering
\includegraphics[width=60mm]{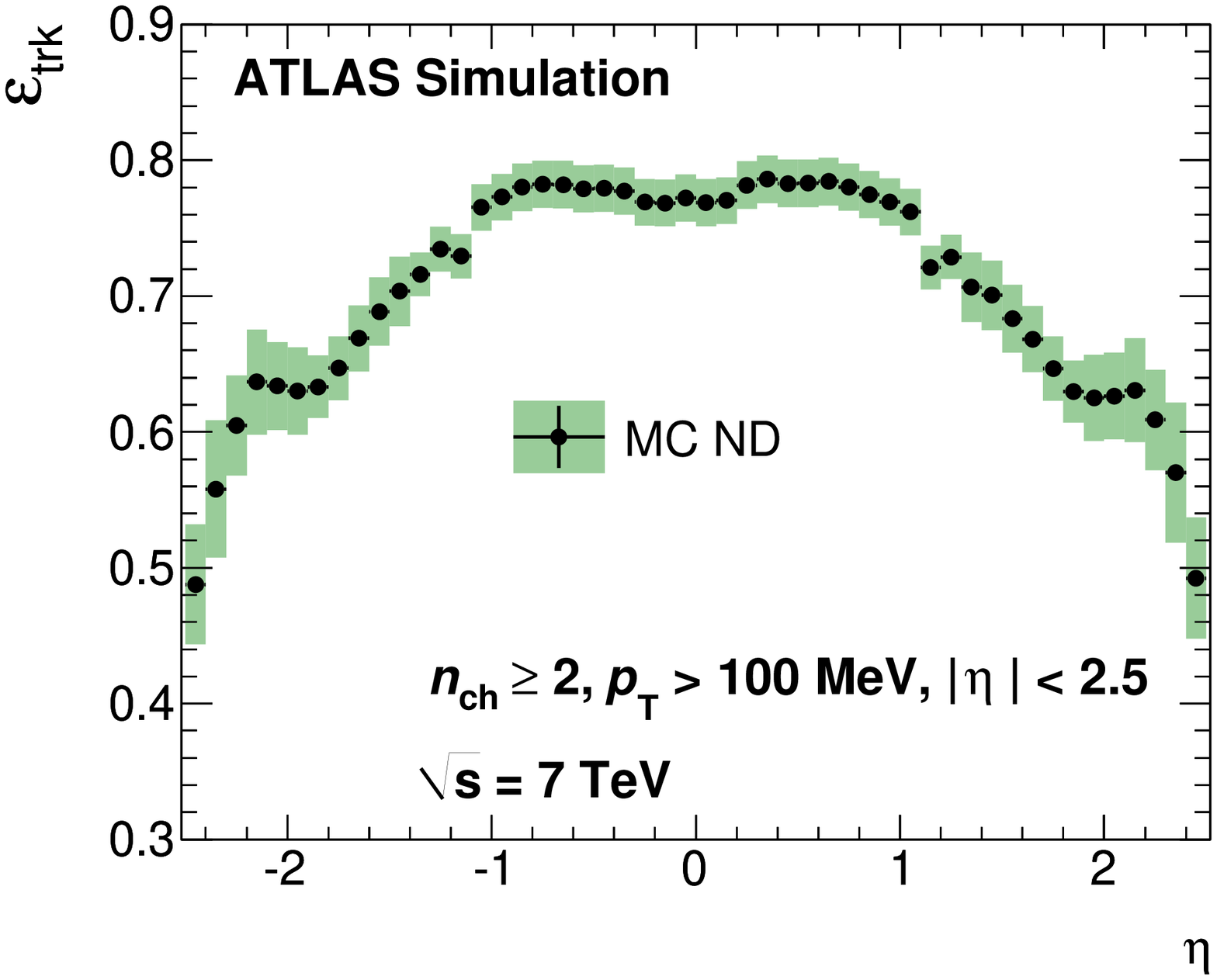}\\
\end{minipage}
\begin{minipage}[b]{60mm}
\centering
\includegraphics[width=60mm]{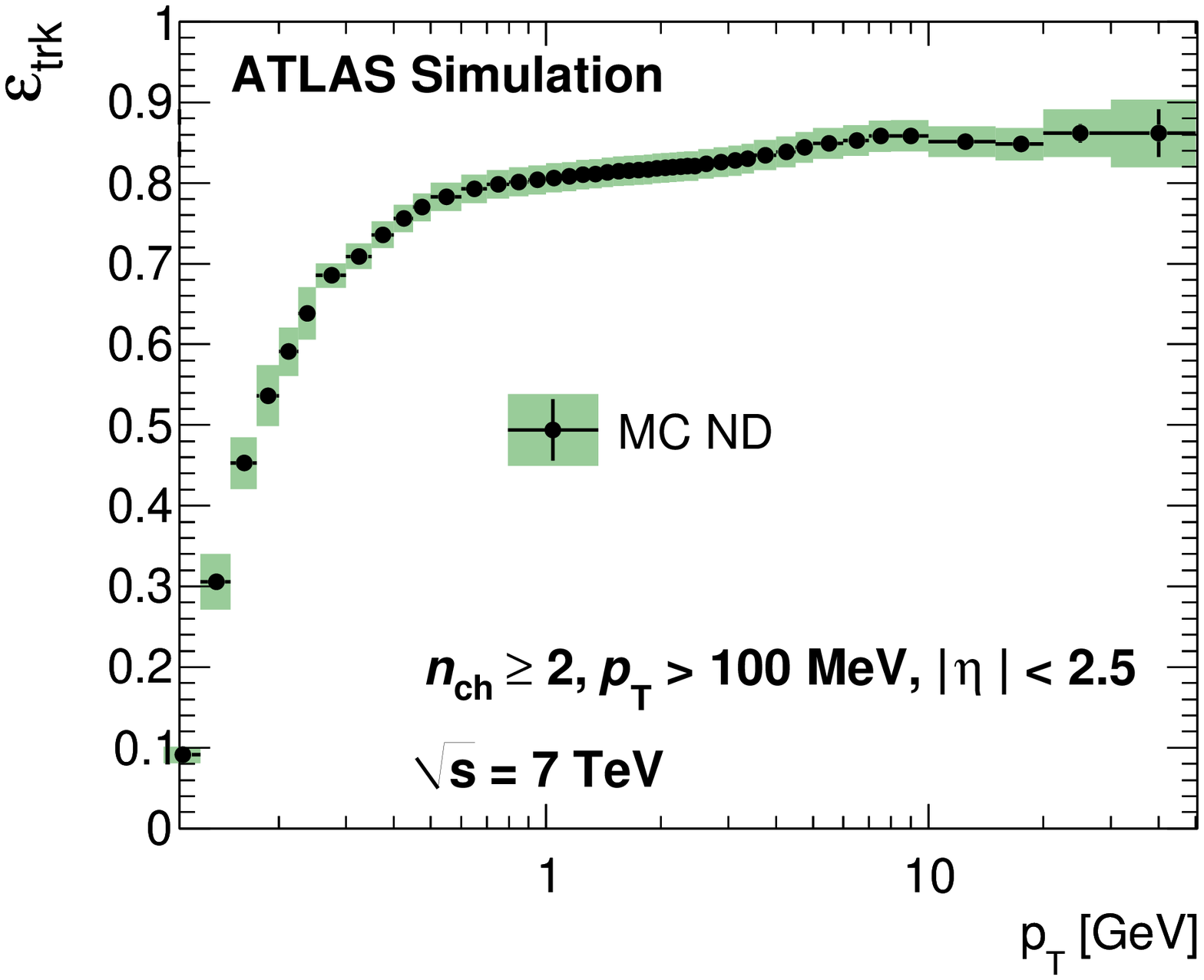}\\
\end{minipage}
  \caption{The track reconstruction efficiency as a function of $\eta$ (left) and $p_{T}$ (right), derived from non-diffractive (ND) Monte Carlo \protect\cite{tracking}.
  \label{fig:track}}
\end{centering}
\end{figure}

Tracks above a given $p_{T}$ threshold (nominally 0.4 GeV, however this value changes depending on the data) are reconstructed offline
within the full acceptance range $|\eta| < 2.5 $ of the Inner Detector, using multi-stage track identification
algorithms. The inside-out algorithm starts from silicon space point seeds  and
adds hits from neighboring silicon layers. The track
candidates found in the silicon detectors are then extrapolated to
include measurements in the TRT. It reconstructs most primary
tracks. The outside-in algorithm starts from segments reconstructed
in the TRT and extends them inwards by adding silicon hits. It
reconstructs secondary tracks (e.g. conversions, hadronic
interactions, $V^{0}$ decays). The track reconstruction efficiency
is defined as the fraction of primary particles with $p_{T} > 100$ MeV and $|\eta| < 2.5 $ matched to a reconstructed track. Figure
\ref{fig:track} presents tracking efficiency for minimum bias
analysis derived from simulation. Efficiency is highest at central
region of rapidity and for tracks with high transverse momentum.

\begin{figure}[h]
\begin{centering}
\begin{minipage}[b]{60mm}
\centering
\includegraphics[width=60mm]{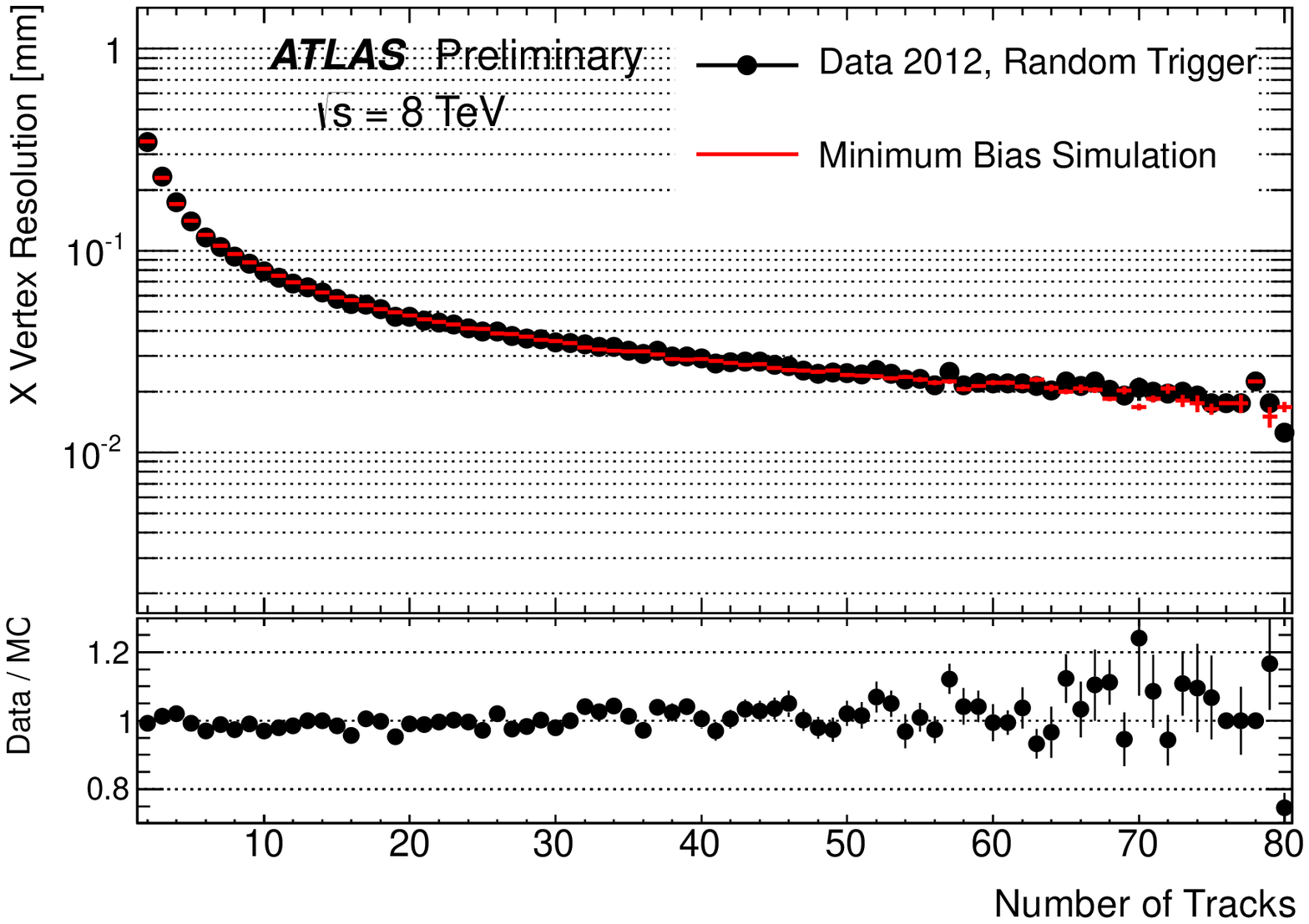}\\
\end{minipage}
\begin{minipage}[b]{60mm}
\centering
\includegraphics[width=60mm]{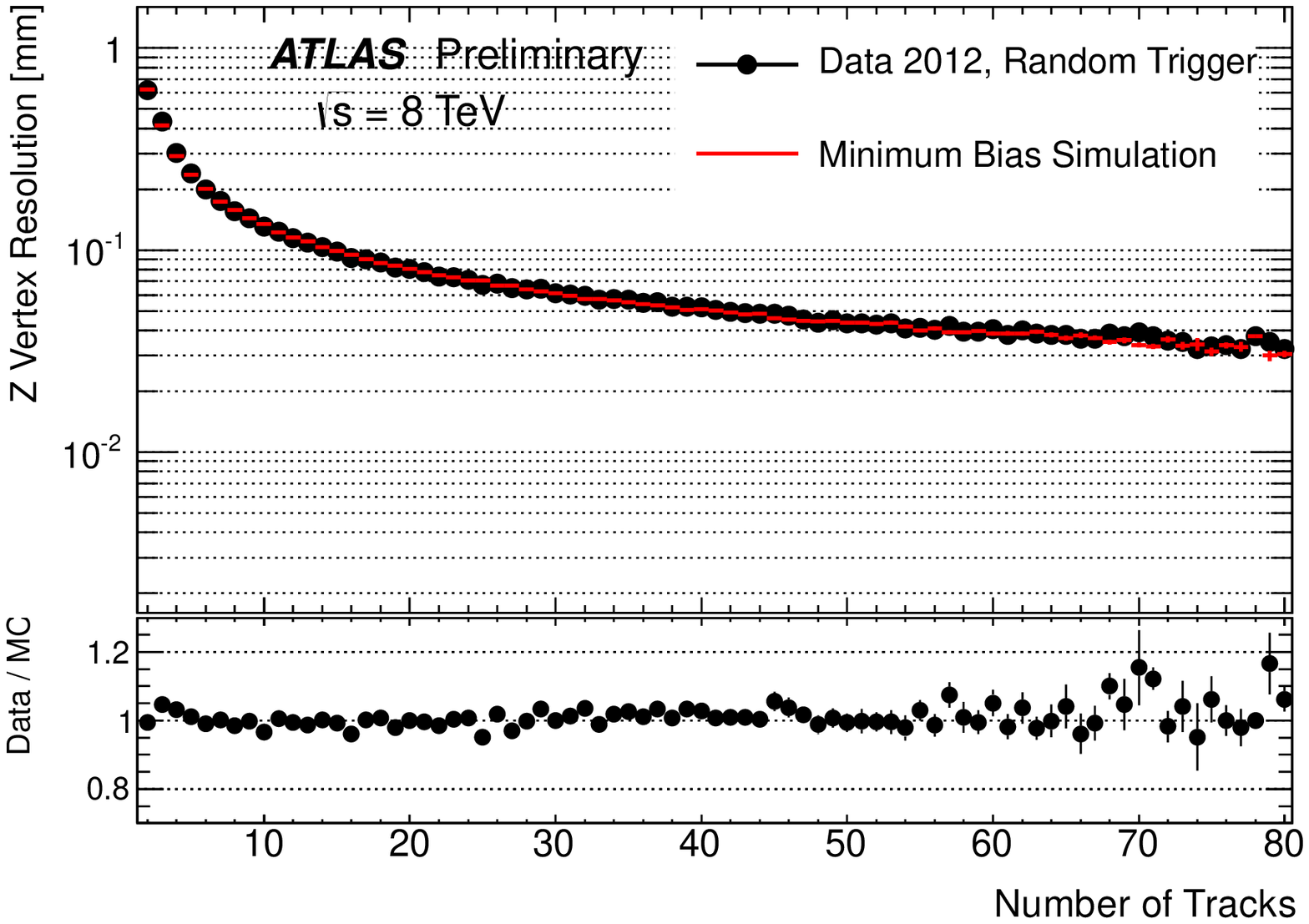}\\
\end{minipage}
  \caption{Vertex position resolution (with no beam constraint) in data (black) and MC (red).
  The resolution is shown for the transverse coordinate (left) and for the longitudinal coordinate (right) as function of the number of tracks in the vertex fit \protect\cite{tracking1}.
  \label{fig:vertex}}
\end{centering}
\end{figure}

Primary vertices are reconstructed using iterative vertex finder
algorithm. Well reconstructed charged tracks which are compatible
with the interaction region are fitted to a common primary vertex.
Then a $\chi^{2}$ is formed and the tracks which are displaced more
than 7 $\sigma$ from the vertex are used to search for the new
vertices. The procedure is repeated until no additional vertices can
be found. The beam spot is used as three-dimensional constraint in
vertex finding algorithm and it is routinely determined from average
vertex position over a short time period. Vertex resolution is
derived from data using split vertex technique. Vertex position
resolution in collision data from 2012 compared with Monte Carlo is
presented on figure \ref{fig:vertex}. There is a very good agreement
at the level of 5\% between data and simulation.

\section{Tracking in High Pile-up}
The luminosity delivered by LHC is the biggest challenge for
tracking and vertexing, as the ID is particularly sensitive to the
increase in particle multiplicity with pile-up. In 2012 the mean
number of inelastic proton-proton interactions ($\mu$) per bunch
crossing reached the number of approximately 30, which is beyond ID
design specification. The increased detector occupancy can result in
degraded track parameter resolution due to incorrect hit assignment,
decreased efficiency and fake tracks from random hit combinations.
This in turn impacts vertex reconstruction, resulting in a lower
efficiency and an increased fake rate. With the data delivered up to
date the track reconstruction efficiency is unchanged. However, when
using the track quality requirements developed for high efficiency
in low pile-up conditions, the fraction of combinatorial fakes
increases with $\mu$ \cite{tracking4}. The fake track fraction in
high pile-up data can be suppressed by applying tighter quality
requirements on reconstructed track, called "robust track
selection". With the robust requirements fake track fraction is
reduced by a factor of 2-5 and becomes almost independent of the
amount of pile-up, but there is moderate drop in primary track
reconstruction efficiency (2-5\%), which is independent of $\mu$
\cite{tracking4}. The probability of reconstructing fake vertices
increases with pile-up. While at low pile-up its value is below
0.1\%, it increases up to 7\% at $\mu$ = 41 when vertices are
reconstructed with tracks passing the default requirements. It can
be controlled with negligible efficiency loss by applying robust
requirements on the tracks used in vertex reconstruction
\cite{tracking4}.

\section{Alignment}
Precise detector alignment is required to obtain ultimate track
parameter resolution. The ID is aligned using a track based method
\cite{align}, which relies on the minimization of a $\chi^{2}$
constructed from track-hit residuals (residual is the difference
between the reconstructed track and the actual hit in the detector).
The alignment is performed at 3 different levels of granularity
corresponding to the mechanical layout of the detector: Level 1
corresponds to entire sub-detector barrel and end-caps of Pixel Detector,
SCT and TRT. Level 2 deals with silicon barrels and discs, TRT
barrel modules and wheels. Level 3 aligns each silicon module and
TRT straws having $\sim$700,000 degrees of freedom in total. Level 1
alignment is performed automatically on run-by-run basis.
The studies on 2011 data show limited movements of Level 1
structures which usually can be correlated to sudden change in
detector conditions, e.g. temperature. There exist systematic
detector deformations e.g. distortion, that cannot be detected using
the approach described above as they retain the helical form of
tracks at the same time biasing the track parameters. Advanced
alignment methods using Z resonance and $E/p$ for electrons are used
to remove residual biases on momentum reconstruction \cite{align1}.

\section{Conclusions} The ATLAS Inner Detector has been fully
operational throughout all LHC data taking periods. The operational
efficiency of all sub-systems is excellent and tracking performance
meets or exceeds design specifications.

\end{document}